# Fractal zone plates for wideband imaging with low chromatic aberration and extended depth of field

**Walter D. Furlan(*) and Genaro Saavedra**

Departamento de Óptica, Universitat de València, E-46100 Burjassot, Spain

(*) walter.furlan@uv.es

**Juan A. Monsoriu**

Departamento de Física Aplicada, Universidad Politécnica de Valencia, E-46022 Valencia, Spain

Fresnel zone plates are diffractive elements that are essential to form images in many scientific and technological areas, especially where refractive optics is not available [1-4]. One of the main shortcomings of Fresnel zone plates, which limit their utility with broadband sources, is their high chromatic aberration. Recently presented, Fractal Zone Plates (FZPs) [5] are diffractive lenses characterized by an extended focal depth with a fractal structure. This behaviour predicts an improved performance of FZPs as image forming devices with an extended depth of field and a reduced chromatic aberration. Here we report the achievement of the first polychromatic images obtained with a FZP that confirm these predictions. We show that the polychromatic modulation transfer function (MTF) of a FZP affected by defocus is about two times better than one corresponding to a Fresnel zone plate. This result opens the possibility to improve imaging techniques that use conventional Fresnel zone plates over a wide range of the electromagnetic spectrum, ranging from x-rays to THz.

Since their introduction in 2003, FZPs have deserved the attention of several research groups working on diffractive optics [6-8], because it has been assumed that they can improve the performance of classical Fresnel zone plates in certain applications where multiple foci are needed. FZPs also inspired the invention of novel photonic structures such as spiral fractal zone plates [8], fractal axicons [9], fractal conical lenses [10], and fractal photon sieves [11]. In spite of their potential applications as imaging elements, up to the present only the focusing properties of FZPs were considered in the literature. Here we examine the incoherent imaging characteristics of these elements using polychromatic light in the visible range. The performance of FZPs is compared with conventional Fresnel zone plates of the same focal distances. We discovered that the FZP provides an extended depth of field and a reduction of chromatic aberration. This finding is confirmed objectively using the modulation transfer function as a merit function.

FZPs can be built using the same methods and technologies used for conventional Fresnel zone plates. Thus, we begin by tracing a parallelism in the design process between these two elements. As it is well known, a Fresnel zone plate consists of alternately transparent and opaque zones whose radii are proportional to the square root of the natural numbers. Thus, it can be generated starting from a 1-D compact-supported periodic function $q(\varsigma)$ [see for example the binary function in Fig. 1(a)] followed by a change of coordinates $\varsigma=(r/a)^2$, and then by rotating the transformed function around one of its extremes. The resulting radially symmetric 2-D structure is a Fresnel zone plate having a radial coordinate $r$ and an outermost ring of external radius $a$ [see Fig. 1(b)]. In a similar way a FZP can be constructed by replacing the periodic function $q(\varsigma)$ by a function $f(\varsigma)$ with fractal profile. In Fig 1 (c) we have represented the 1-D binary function associated to a member of the triadic Cantor set developed up to the third stage, S=3 [5]. The corresponding FZP is shown in Fig 1 (d). It is worth to mention that in the most general case any 1-D fractal can be employed to construct a FZP.

A comparison between Fig. 1(a) and Fig. 1(c) shows that the minimum distance between the zones of the fractal binary function $f(\varsigma)$ coincides with half period of the periodic function $q(\varsigma)$. This distance defines the width of the outermost zones in the zone plates, $\Delta r$, and thus limits their ultimate spatial resolution. However, as the focal length of a Fresnel zone plate is inversely proportional to the wavelength, when using wideband illumination, its longitudinal chromatic aberration [see Fig. 2(a)] severely limits its imaging performance. We will see that FZPs can partially overcome this limitation.

We calculate the monochromatic irradiances along the optical axis provided by the zone plates using the Fresnel diffraction formula:

$$I(z) = \left(\frac{2\pi}{\lambda z}\right)^2 \left| \int_0^a p(r) \exp\left(-i\frac{\pi}{\lambda z} r^2\right) r \, dr \right|^2, \qquad (1)$$

where $p(r)$ is the 2-D circularly symmetric pupil function that describes the zone plate in canonical polar coordinates. In Fig. 2 we represented the axial irradiances provided by a Fresnel and a fractal zone plates of the same focal length computed for three different wavelengths in the visible spectrum. As can be seen, the principal lobes of the different foci provided by a FZP coincide in the axial position with those obtained with a conventional Fresnel zone plate, but the FZP produce a sequence of subsidiary foci around each major focus following the fractal structure of the FZP itself. In general, the position, the intensity and the 3-D shape of the focal spot depend on the fractal level and on the lacunarity [12] of the encoded fractal function $f(\varsigma)$. In all cases these subsidiary foci of the FZP provide an extended depth of focus for each wavelength that partially overlap with the other ones creating an overall extended depth of focus which is less sensitive to the chromatic aberration.

We have experimentally tested the imaging capabilities of FZPs under white light illumination and compared their performance against a conventional Fresnel zone plate of the same main focal distance. In our experiment we used a Fresnel zone plate with 81 zones and the equivalent FZP constructed using a triadic Cantor set developed up to stage S=4 [i.e. doubling the number of zones of the FZP in Fig.1(d)]. The diffractive lenses were printed and then photographically reduced onto 35mm slides resulting with a radius $a$=2.52 mm and focal distance $f_{ZP}(\lambda_o) = 124$ mm for the design wavelength $\lambda_o = 633$ nm. In the experimental setup, a conventional *object-lens-image plane* arrangement, the test object consisted of a set of binary letters from an optotype-like chart, was illuminated using a polychromatic light source (Fiber optic illuminator, with a 150 W EKE 3220 K halogen lamp). The images were obtained directly onto the CMOS detector of a Cannon EOS 350D digital camera with 6.4 μm square pixel size. In the experiment, the distance $L$ between the object and the image plane was fixed to 575 mm and, to obtain the out of focus images, the distance $d$ from the object plane to the zone plates was varied. Accordingly, the defocus coefficient that affected the image at the output plane can be expressed as:

$$w(d,\lambda) = -\frac{a^2}{2\lambda}\left(\frac{L}{d(L-d)} - \frac{1}{f_{ZP}(\lambda)}\right) \qquad (2)$$

The results are summarized in Fig. 3. Due to the different transmittances of both kind of lenses the range of intensities of the photographs in this figure were normalized to the peak intensity, but no additional post-processing was performed. Defocused images obtained with the FZP show clear improvements when compared to those with Fresnel zone plate. The improvement is particularly noticeable with respect to the chromatic blur which is considerable lower in the images obtained with FZP, even at the focal plane (however at this plane FZP provides a slightly poor resolution). As a result of the less sensitivity of FZP to the chromatic aberration the depth of field achieved with this plate is larger than the one obtained with Fresnel zone plate. In order to confirm the result of a subjective comparison, we compute the MTF values for the two lenses in the three experimental conditions. This computational modeling, shown in Fig. 4, accounts for the wideband illumination (wavelengths between 488 and 647 nm) employed in the experiment. The polychromatic MTFs were computed as the superposition of three monochromatic MTF (using the same wavelengths as in Fig. 2) weighted by the spectral content of the illumination source and the spectral response of the detector. The results are plotted in Fig. 4. We found that the MTFs for the defocused planes is about 2 times better for FZP. Concretely, for the range of frequencies represented in Fig. 4 the ratio between the FZP and Fresnel zone plate MTFs has the following mean values: 2.41 for $d$=345 mm (defocused plane away from the lens); 0.46 at the focal plane; and 1.73 for $d$=440 mm (defocused plane toward the lens). The improved performance of FZPs for defocused planes is particularly evident at medium-high frequencies.

We believe that the improved imaging capabilities (i.e. an increased in the depth of field and a reduction in the chromatic aberration), shown in Fig. 3 and supported by Fig. 4, that FZP offer in polychromatic imaging could be profited across a broad range of applications where conventional Fresnel zone plates are currently applied such as X-ray microscopy (where narrow-band sources are hardly available); THz imaging and tomography (to project line spots onto the object efficiently [13]), astronomy [14] and ophthalmology (in the form of intraocular or contact lenses). Moreover, because of the fact that FZP can be produced using the same techniques to those for making Fresnel zone plates all the improvements already reported for them, such as increasing the resolution by the overlay nanofabrication technique [15] or by the use of composite zones [16]; are still valid for FZP fabrication. Interestingly, in FZP design there is an additional parameter, the lacunarity of the fractal function, that adds versatility to the design process because it can modify the number of foci and their relative amplitude [12].

**Acknowledgments**

This research has been supported by the following grants: 1) DPI 2006-8309, Plan Nacional I+D+I, Ministerio de Ciencia y Tecnología. Spain. 2) "Programa de Incentivo a la Investigación UPV 2005", Vicerrectorado de Innovación y Desarrollo, Universidad Politécnica de Valencia, Spain

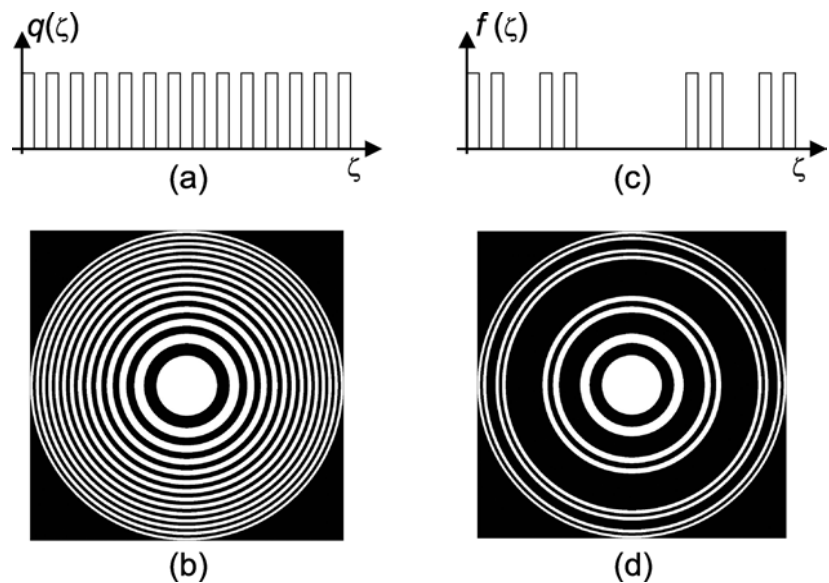

Figure 1. **An illustration of the fractal zone plates design**. The same procedure can be followed in designing Fractal and Fresnel zone plates. It consists in 3 steps. (**a**) Fresnel zone plate starts from a binary 1-D compact supported function $q(\zeta)$. After the change of variables $\zeta=(r/a)^2$. The resulting function, which is periodic in the new coordinate, is rotated around the origin. The result is the zone plate represented in **b.** (**c**) The originating function for the FZP is a fractal binary 1-D function $f(\zeta)$. In this case it is a triadic Cantor set for stage 3 (see Ref. 5 for details). (**d**) The change of variables and rotation performed to obtain **b** now results in a FZP.

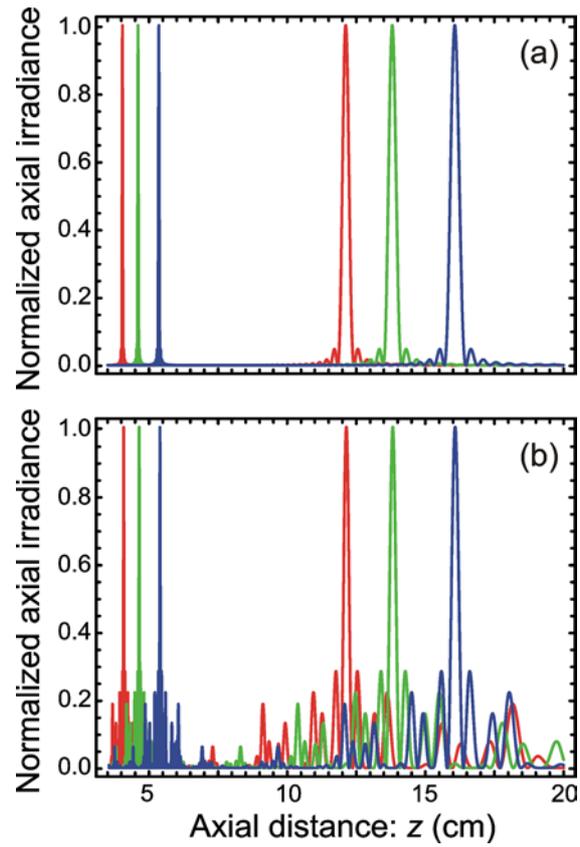

Figure 2. **Point spread functions for different wavelengths**. The axial irradiances were computed using equation (1) for *a*=2.52 mm and λ=647nm (red line), λ=568nm (green line), and λ=488nm (blue line). (**a**) Fresnel zone plate of 81 zones. (**b**) FZP of the same focal distance. The FZP in this case was obtained for a triadic Cantor set like the one in Fig. 1 (d) but developed up to stage 4.

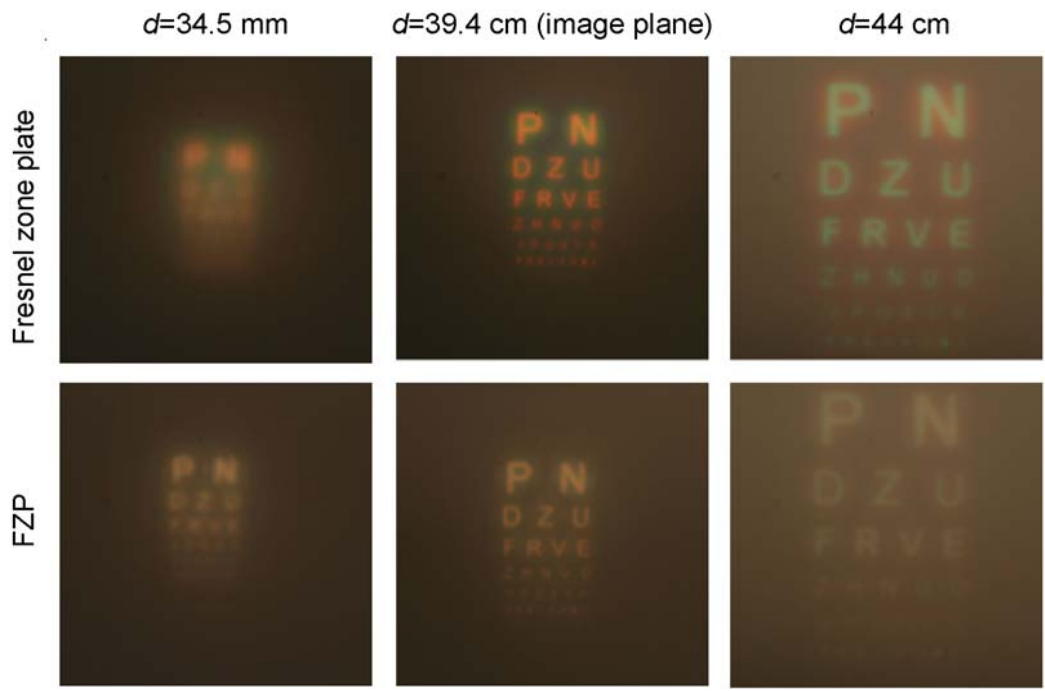

Figure 3. **Images obtained with FPS and with Fresnel zone plate**. Three different locations for the image plane were considered: In front of the paraxial image plane (*d*=44.0 cm), at the paraxial image plane (*d*=39.4 cm), and behind of the paraxial image plane (*d*=34.5 cm).

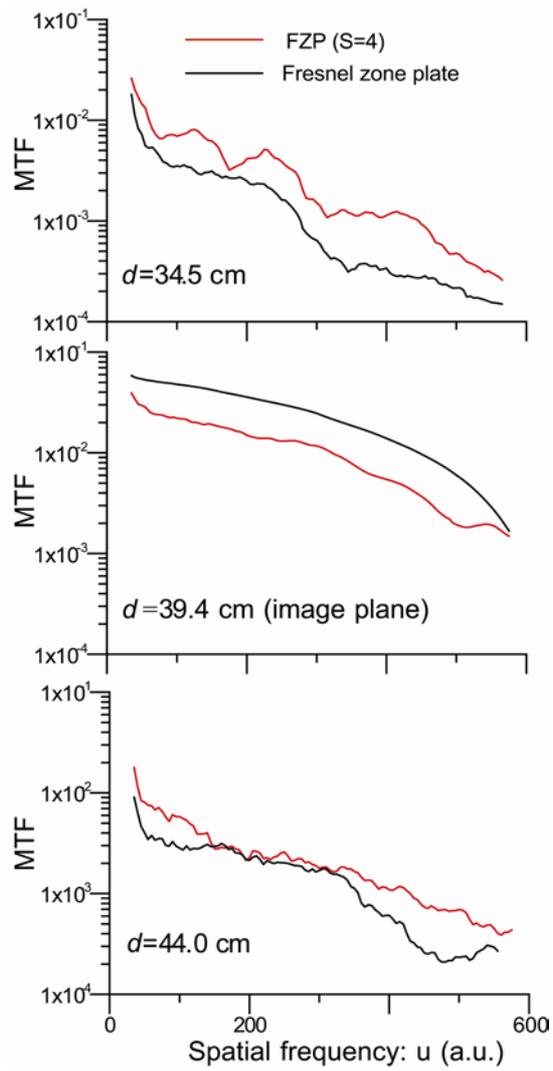

Figure 4. **Modulation transfer functions for defocused planes.** The MTFs correspond to the three locations of the image planes in Fig. 3.